\documentclass[useAMS,usenatbib]{mn2e}
\usepackage[figuresright]{rotating}
\usepackage{rotating}
\usepackage{color}
\usepackage{longtable}
\usepackage{lscape}
\usepackage{lipsum}
\usepackage{lscape}
\usepackage{xspace}

\title[AGN abundances comparison]{Chemical abundances of  Seyfert 2 AGNs-- I.   
Comparing oxygen abundances from distinct methods using SDSS}

\author[Dors et al.]
  {O.~L. Dors$^{1}$\thanks{E-mail: olidors@univap.br}, P. Freitas-Lemes$^{1}$, E.~B.~Am\^{o}res$^2$,
  E. P{\'e}rez-Montero$^3$, M.~V. Cardaci$^{4,5}$,  \newauthor{G.~F. H\"agele$^{4,5}$, M. Armah$^{1}$, 
A.~C. Krabbe$^{1}$,  M. Fa{\'u}ndez-Abans$^6$}\\
$^1$ Universidade do Vale do Para{\'i}ba. Av. Shishima Hifumi, 2911, CEP:
12244-000, S{\~ a}o Jos{\'e} dos Campos, SP, Brazil\\
$^2$ Departamento de F{\'i}sica, Universidade Estadual de Feira de Santana, Av. Transnordestina, S/N, CEP 44036-900 Feira de Santana, BA, Brazil \\
$^3$ Instituto de Astrof{\'i}sica de Andaluc{\'i}a, Camino Bajo de Hu{\'e}tor s/n, Aptdo. 3004, E18080-Granada, Spain.\\
$^4$ Instituto de Astrof\'{\i}sica de La Plata, Argentina\\
$^5$ Facultad de Ciencias Astron\'omicas y Geof\'{\i}sicas, Universidad Nacional de La Plata, Paseo del Bosque s/n, 1900 La Plata, Argentina.\\
$^6$ MCTIC/ Laborat{\'o}rio Nacional de Astrof{\'i}sica, CEP:37.504-364, Itajub{\'a}, MG, Brazil.\\
}
  
\date{Released 2019 Apr 15}

\def\cm2{cm$^2$ }
\def\se1{s$^{-1}$ }

\def\arcsec{\hbox{$^{\prime\prime}$} }

\definecolor{green2}{rgb}{0,0.7,0}

\newcommand{\ion}[2]{{\textrm{#1}}\,{\textrm{\sc #2}}}

\begin{document}


\maketitle
\begin{abstract}
We compare the oxygen abundance (O/H) of the Narrow Line Regions (NLRs) of  Seyfert 2 AGNs obtained  through
strong-line methods and from direct measurements of the electron temperature ($T_{\rm e}$-method). The aim of this
study is to explore the effects of the use of distinct methods on the range of metallicity
and on the mass-metallicity relation of AGNs at low redshifts ($z \: \la \: 0.4$). 
 We used the  Sloan Digital Sky Survey (SDSS) and NASA/IPAC Extragalactic Database (NED) to selected
optical ($3000 \: < \: \lambda({\mathrm{\AA}}) \: < \: 7000$) emission line intensities of 463 confirmed  
Seyfert 2 AGNs. The oxygen abundance of the NLRs were estimated using the theoretical Storchi-Bergmann et al. calibrations, the semi-empirical
$N2O2$ calibration, the bayesian {\sc \ion{H}{ii}-Chi-mistry}   code and the $T_{\rm e}$-method. 
 We found that the oxygen abundance estimations via the strong-line methods differ from each other
up to $\sim0.8$ dex,   with the largest discrepancies in the low metallicity regime 
($\rm 12+\log(O/H) \: \la \: 8.5$).  We confirmed that the 
$T_{\rm e}$-method underestimates the oxygen abundance in NLRs,  producing unreal subsolar values.
 We did not find any correlation between the stellar mass of the host galaxies and the
metallicity of their AGNs. This result is independent of the method used to estimate $Z$.    
\end{abstract}


\section{Introduction}

 Active Galactic Nuclei (AGNs) and Star-Forming regions (SFs) 
present in their spectra  prominent emission-lines  whose relative intensities can be used to estimate
the chemical abundances of the heavy elements  in the gas-phase of these objects at a wide redshift range.
Therefore, metallicity estimations in AGNs and in SFs are essential   in  
 the study of galaxy formation and chemical evolution of the Universe.

Along  decades, metallicity ($Z$) and  relative abundance  of heavy elements   (e.g. N/O, C/O) have been estimated in a large sample of
SFs at low and high redshifts (see \citealt{2019A&ARv..27....3M} for a review).
 There is a  consensus   that a reliable estimation of $Z$  can be obtained with 
 a previous   direct  measurement of the electron temperature of the gas, i.e. by the $T_{\rm e}$-method 
(e.g. \citealt{2003ApJ...591..801K, 2006MNRAS.372..293H, 2008MNRAS.383..209H}).
The use of the $T_{\rm e}$-method requires to measure temperature-sensitive line ratios, such
as [\ion{O}{iii}]($\lambda$5007/$\lambda$4363), but the [\ion{O}{iii}]$\lambda$4363
is too   weak or unobservable in several SFs with high $Z$ and/or low ionization degree \citep{2002MNRAS.329..315C, 2013ApJ...774...62L, 2007MNRAS.375..685P, 
2007MNRAS.382..251D, 2008A&A...482...59D, 2009MNRAS.398..485P}. For such objects, along decades, calibrations between 
$Z$  and more easily  measurable line-ratios, defined as strong-line methods  \citep{1979MNRAS.189...95P}, 
have been  suggested by several   authors  (see \citealt{2010A&A...517A..85L} for a review).
 The main problem associated with metallicity estimations of SF is that $Z$ values obtained  using
the $T_{\rm e}$-method and those based on theoretical strong-line methods are not in agreement, in the sense that the former 
method produces $Z$ values lower (by about 0.2 dex)
than those from the latter \citep{2003ApJ...591..801K, 2005A&A...437..837D, 2010A&A...517A..85L, 2011MNRAS.415.3616D}.
  This problem is called ``temperature problem'' and its origin  is an open problem in the nebular astrophysics.

  Contrary to SFs, metallicity determinations in AGNs have received little attention. 
 In fact, the first quantitative abundance determinations for the O/H and N/H and  for a large sample of AGNs (Seyfert 2) seems to be  the 
one performed by \citet{2017MNRAS.468L.113D}, who  built detailed
photoionization models to reproduce narrow optical emission-line intensities of a
 sample of 47 objects   (see also \citealt{2015MNRAS.453.4102D, 2019MNRAS.486.5853D}).
 Thereafter, \citet{thomas18a} and \citet{2018ApJ...867...88R}  also carried out oxygen abundance
estimations for a few Narrow Line Regions (NLRs) of AGNs (see also  \citealt{1992A&A...266..117A, 2014MNRAS.439..771B, 2011ApJS..197...32W, 2007ApJ...658..804D, 2003ApJ...583..649B,  
2002ApJ...564..592H,  1996ApJ...461..683F, 1993ApJ...418...11H, 1992ApJ...391L..53H, 2018ApJ...856...46R}).
Moreover, few works have been done to develop
 methodologies to estimate $Z$ in AGNs.  Currently, there are only four calibrations between the metallicity and  
 narrow strong emission-lines of AGNs proposed by \citet{1998AJ....115..909S}, \citet{2014MNRAS.443.1291D, 2019MNRAS.486.5853D} and \citet{2017MNRAS.467.1507C}
 and  three  Bayesian methods proposed by \citet{thomas18a},  \citet{2019A&A...626A...9M} and \citet{2019MNRAS.489.2652P} in the literature.
 It is worth to mention that, the level of metallicity discrepancies derived from distinct AGN calibrations  
  have been investigated considering only few objects \citep{2015MNRAS.453.4102D, 2017MNRAS.467.1507C, 2018ApJ...856...46R}.
  Specifically,  \citet{2015MNRAS.453.4102D} showed the exis\-tence of the  temperature problem  in AGNs but, these authors 
used a few number (for 44 Seyfert 2 nuclei) of abundance estimations.

Another important point is  the observational database. Recent surveys, such as 
the Calar Alto Legacy Integral Field Area (CALIFA) survey \citep{2012A&A...546A...2S} and the Sloan Digital Sky Survey \citep[SDSS;][]{2000AJ....120.1579Y},  
have  produced a very large sample of spectroscopic database and the use of these data have
revolutionized the extragalactic astronomy. 
However, the observational data from these surveys have been mostly  used
for the study of the chemical abundances in SFs (e.g. \citealt{2004ApJ...613..898T, 2006A&A...453..487S, 2006A&A...459...85N, 2006ApJ...652..257L, 
2016A&A...595A..62P, 2016MNRAS.462.2715Z, 2017MNRAS.469.2121S, 
2019A&A...624A..21G, 2016ApJ...819..110S, 2013MNRAS.432.1217P, 2008ApJ...681.1183K}) 
 while the $Z$ determination in AGNs  has  been barely explored.
In fact, \citet{2012MNRAS.427.1266V} used the SDSS-DR7 data  \citep{2009ApJS..182..543A} to derive
the internal reddening, ionization parameter, electron temperature, and
electron  density of about 2100 Seyfert 2 galaxies but the oxygen abundance or metallicity were not
estimated in this analysis. \citet{2013MNRAS.430.2605Z} also
used the SDSS data to determine the electron density and  electron temperature   of active 
and star-forming nuclei. These authors did not produce additional estimations of the metallicity
for the considered sample  (see also \citealt{2014MNRAS.437.2376R,  2009MNRAS.397..172G}).

With the above in mind, the emission-line 
intensities of the SDSS-DR7 \citep{2009ApJS..182..543A} measured 
by the MPA-JHU\footnote{Max-Planck-Institute for Astrophysics and John Hopkins University} group  are used in this paper in order to calculate the oxygen
abundances for a large number of Seyfert 2s, whose classifications were taken from the
NASA/IPAC Extragalactic Database (NED).
Our main  goals are:
\begin{itemize}
\item   Making available emission-line intensities of a large sample of Seyfert 2  AGNs.
\item   Comparing the oxygen abundances of Seyfert 2 AGNs obtained using different methods.
\item   Investigating the effect of the use of distinct methods on the mass-metallicity relation.
\end{itemize}

The present study is organized as  it follows. In Section~\ref{observ},
a description of the observational data and a discussion  about aperture effects are presented. In Section~\ref{oxy}, the methodology
used to estimate the oxygen abundance and other parameters of the sample are presented. The results 
and discussion are given in Sects.~\ref{res} and \ref{disc}, respectively. The conclusion of the outcome is presented 
in Sect.~\ref{conc}.

\section{OBSERVATIONAL SAMPLE}
\label{observ}

\begin{figure*}
\centering
\includegraphics[angle=+90,width=1\columnwidth]{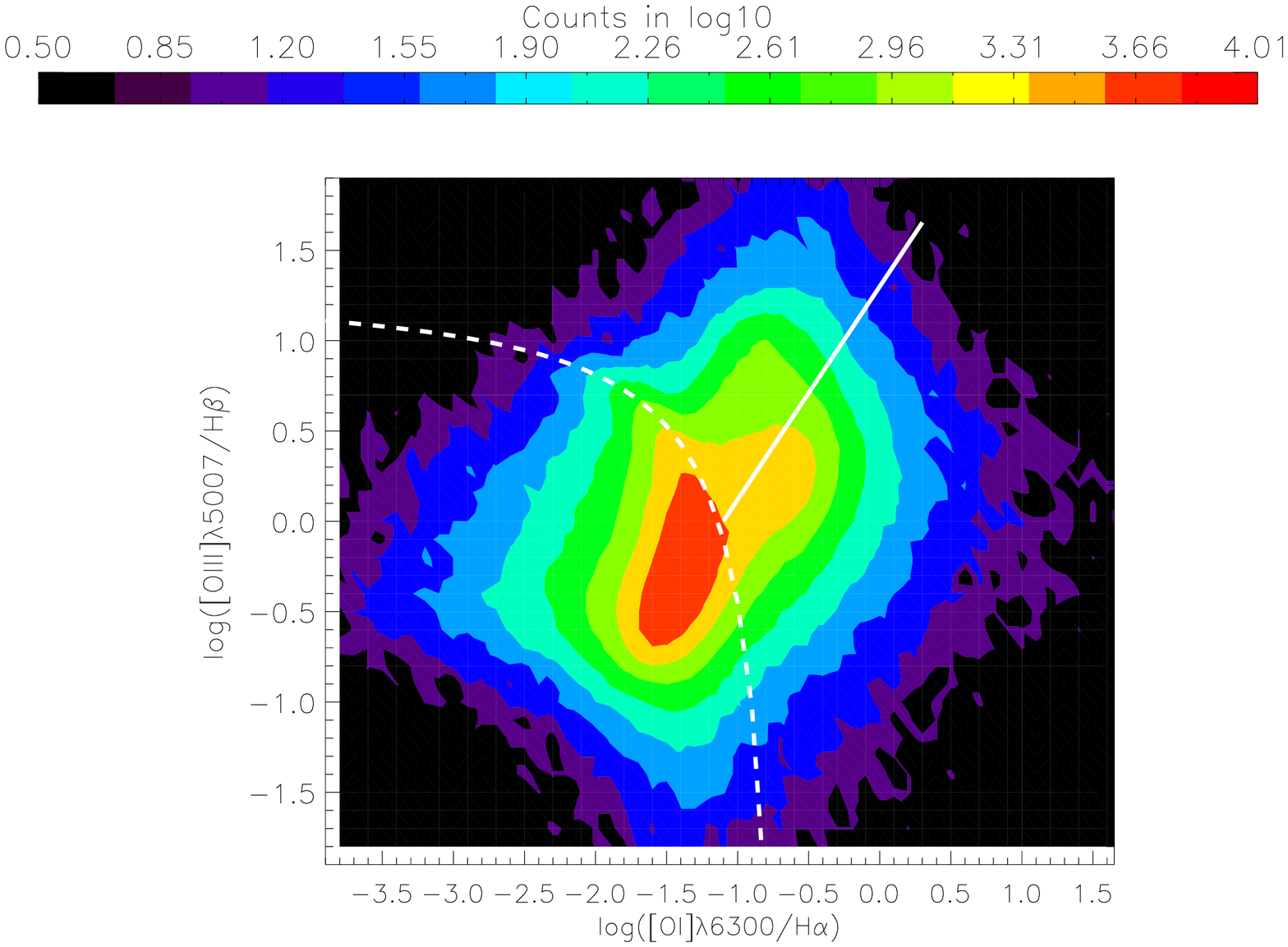}
\includegraphics[angle=+90,width=1\columnwidth]{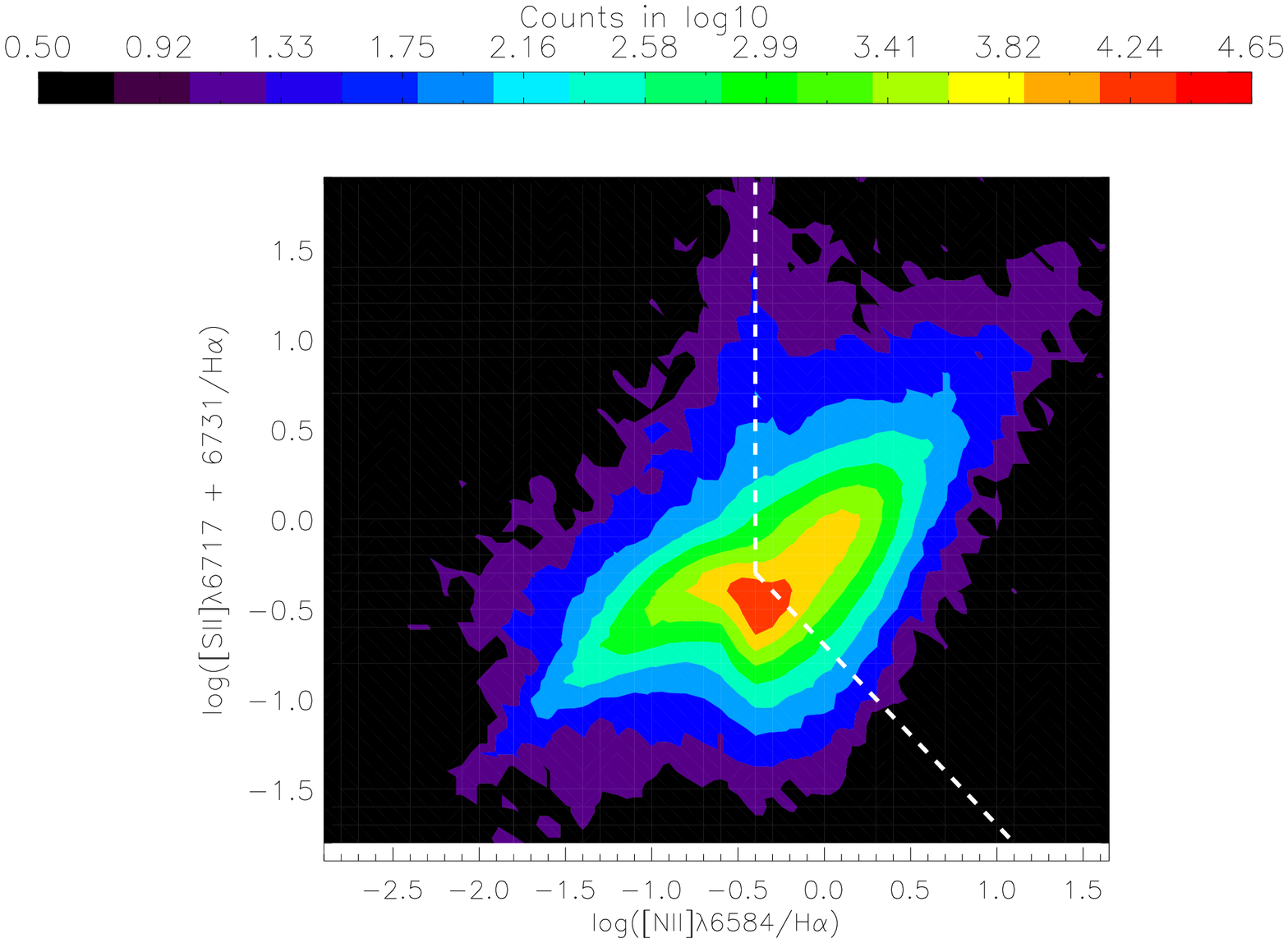}
\includegraphics[angle=+90,width=1\columnwidth]{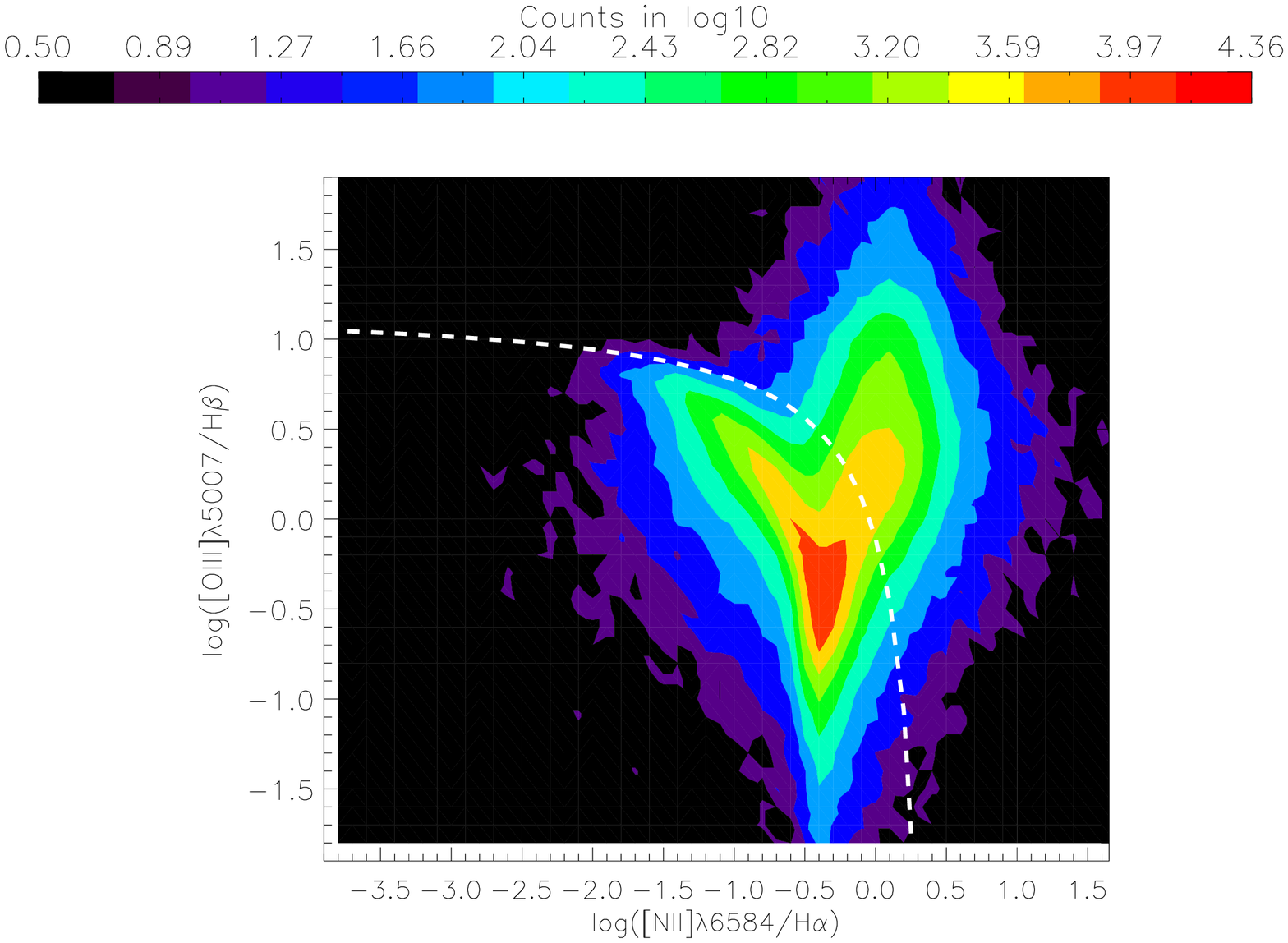}
\includegraphics[angle=+90,width=1\columnwidth]{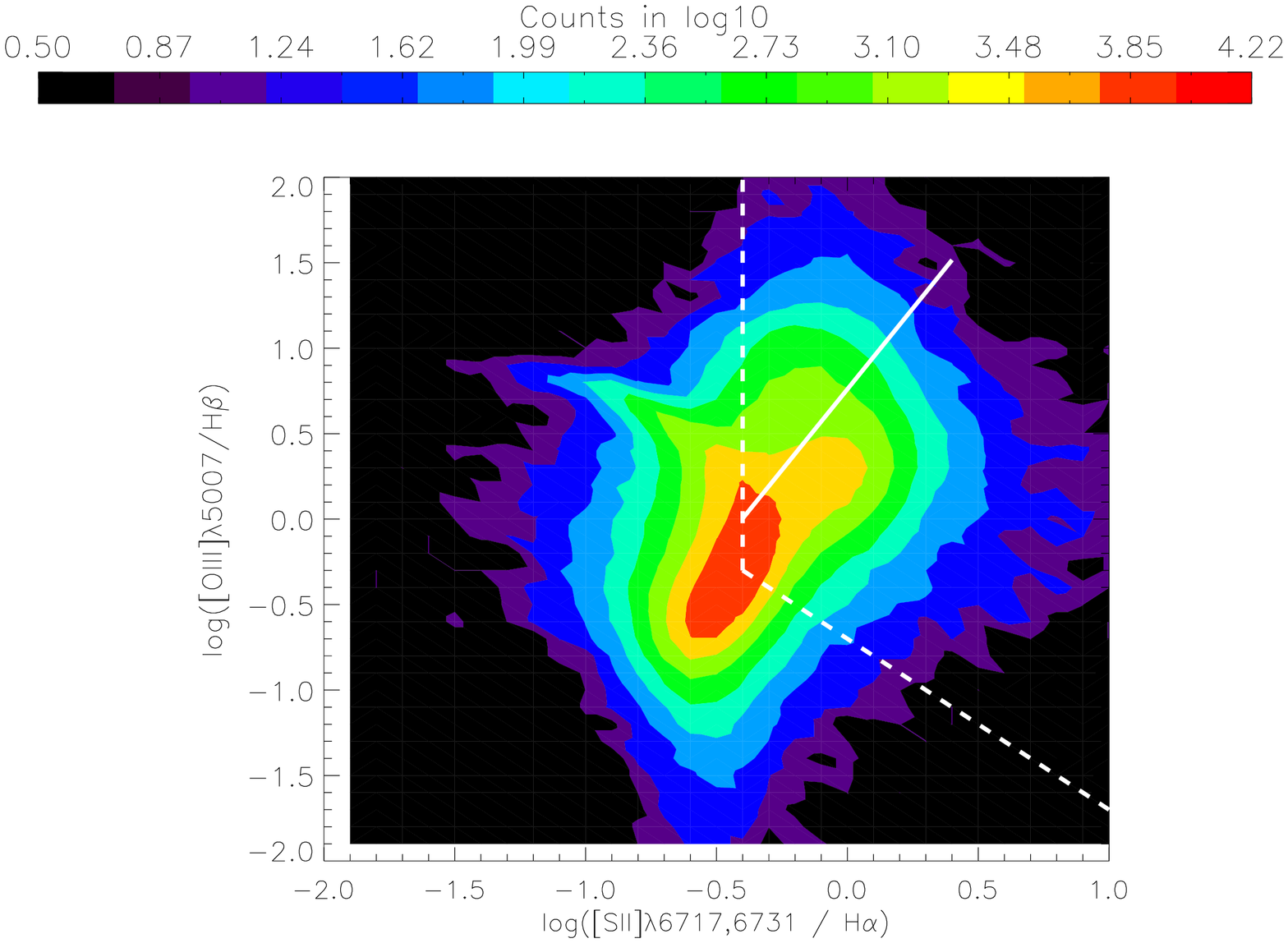}
\caption{Diagnostic diagrams log([\ion{O}{iii}]$\lambda$5007/H$\beta$) versus log([\ion{O}{i}]$\lambda$6300/H$\alpha$,
log([\ion{S}{ii}]$\lambda\lambda$6716+31/H$\alpha$) versus log[\ion{N}{ii}]$\lambda$6584/H$\alpha$, 
log([\ion{O}{iii}]$\lambda$5007/H$\beta$) versus log[\ion{N}{ii}]$\lambda$6584/H$\alpha$ and 
log([\ion{O}{iii}]$\lambda$5007/H$\beta$) versus log([\ion{S}{ii}]$\lambda\lambda$6716+31/H$\alpha$).
The  logarithm of the number of compiled SDSS objects (see Sect.~\ref{observ}) are shown according
 to their positions in each panel. }
\label{fig1}
\end{figure*}

\subsection{Observational data}

 In order to produce a sample of type 2 AGNs with observational intensities of  narrow optical emission-lines, we 
used the  measurements of the  Sloan Digital Sky Survey (SDSS, \citealt{2000AJ....120.1579Y}) DR7 
data made available\footnote{https://wwwmpa.mpa-garching.mpg.de/SDSS/DR7/} by MPA/JHU group.
The procedure of measuring the emission-line intensities is described in details by  \citet{2004ApJ...613..898T}.
The data produced by MPA/JHU are corrected for foreground (galactic) reddening using the 
the methodology presented by \citet{1994ApJ...422..158O}.

 In  the SDSS-DR7 database there are 927\,552 objects 
with  signal-to-noise ratio (S/N) larger than 2 and redshift $z \: < 0.7$, in which
 778\,695 objects of these have estimation of stellar mass.
 In order to keep up the consistency of our analysis  with our previous works  (e.g. \citealt{2019MNRAS.489.2652P, 2015MNRAS.453.4102D}), 
we selected only the objects which have, at least, 
the [\ion{O}{ii}]$\lambda3727$, H$\beta$, [\ion{O}{iii}]$\lambda5007$,  [\ion{O}{i}]$\lambda6300$,
H$\alpha$, [\ion{N}{ii}]$\lambda6584$, and [\ion{S}{ii}]$\lambda \lambda$6717,31 emission-lines measured.  By adopting this procedure,
our sample was reduced to 538\,878 objects, mainly due to  the requirement  of having the [\ion{O}{ii}]$\lambda3727$ line measured.

Subsequently,  in order to classify objects as AGN-like and as \ion{H}{ii}-like,  we used 
the standard Baldwin-Phillips-Terlevich (BPT) diagrams \citep{1981PASP...93....5B, 1987ApJS...63..295V}. We used the  criteria
 proposed by \citet{2001ApJ...556..121K} and \citet{2013A&A...549A..25P},  which states that AGN-like are the ones that satisfy:
\begin{equation} \scriptstyle
\rm log([O\:III]\lambda5007/H\beta) \: > \: \frac{0.61}{log([N\:II]\lambda6584/H\alpha)-0.47}+1.19,
\end{equation}

\begin{equation} \scriptstyle
\rm log([O\:III]\lambda5007/H\beta) \: > \: \frac{0.72}{log([S\:II]\lambda\lambda6717+ 31/H\alpha)-0.32} +1.30,
\end{equation}

\begin{equation} \scriptstyle
\rm log([O\:III]\lambda5007/H\beta) \: > \: \frac{0.73}{log([O\:I]\lambda6300/H\alpha)+0.59}+1.33,
\end{equation}
\noindent and
\begin{equation} \scriptstyle
\rm log([N\:II]\lambda6584/H\alpha) \: > \:  -1.05 \: \times \: log([S\:II]\lambda\lambda6717+ 31/H\alpha).
\end{equation}
The ``composite'' objects as defined in  \citet{2006MNRAS.372..961K} are not included in the sample.

In Figure~\ref{fig1}, we present the diagnostic diagrams for the selected  galaxies  from the SDSS-DR7
\citep{2009ApJS..182..543A} with the number of objects in each region according  to the above criteria.
These panels  thus show the known results based on this sample,
according  to the SDSS-DR7 data, there is 
a larger number of  \ion{H}{ii}-like objects than AGN-like  ones (e.g. \citealt{2004MNRAS.351.1151B, 2013MNRAS.430.2605Z}).
We applied the criterion (also shown in Fig.~\ref{fig1}) proposed by \citet{2006MNRAS.372..961K} to the selected sample to separate 
AGN-like and  Low-ionization nuclear emission-line region (LINER) objects. 
The criterion establishes that objects with
\begin{equation} \scriptstyle
\rm log([O\:III]\lambda5007/H\beta) \: < \:  1.30+1.18\: \times \: log([O\:I]\lambda6300/H\alpha)
\end{equation}
\noindent and
\begin{equation} \scriptstyle
\rm log([O\:III]\lambda5007/H\beta) \: < \:  0.76+1.89\: \times \:  log([S\:II]\lambda\lambda6717+ 31/H\alpha)
\end{equation}
\noindent are candidates to be AGN-like objects (including,   for instance, \ Seyfert 1s, Seyfert 2s, quasars, 
\ion{H}{ii}-like objects with very strong winds and shocks), otherwise  they are candidates to be LINERs.  

 As discussed above, the main interest in this paper is to address
the study of AGNs. For this reason, we selected all objects
that appear simultaneously above the dashed lines in the four panels
of Figure~\ref{fig1}.  In total, there are 69,517 objects that
 satisfying the criteria  presented by \cite{2001ApJ...556..121K, 2006MNRAS.372..961K} and
\cite{2013A&A...549A..25P}.

The classification criteria for separating objects according to their main ionization mechanisms 
presented previously and based on BPT diagrams  are defined for objects at redshifts $z\sim 0$. However, 
\citet{2013ApJ...774L..10K} showed that the  demarcation lines in  optical
diagnostic diagrams  change as a function of cosmic time, since interstellar medium conditions are more extreme
and it is expected harder ionizing radiation from stellar clusters (ionizing source of \ion{H}{ii}-like objects) at high redshifts
 than  those in local galaxies. Nevertheless,  such as pointed by  these
 authors, galaxy properties 
practically do not change for $z \: < \: 1$. The maximmum value of the redshift for the objects in our sample is $\sim 0.37$. Therefore,
the cosmic evolution does not influence our classification.

For the selected objects,   all emission line-fluxes  were divided 
by  the corresponding  H$\beta$ flux.  Next, we compiled from the NED/IPAC\footnote{ned.ipac.caltech.edu}
(NASA/IPAC Extragalactic Database) two catalogues containing basic information (classification) about Seyfert 1 and Seyfert 2 galaxies. 
 In total, there are 10,054   classified as  Seyfert 1 and 4,258 as Seyfert 2 AGNs. As the NED/IPAC 
provides a name of SDSS and the Garching's database the \emph{objID},  we matched them using the field \emph{objID} supplied in
both databases. We use the SDSS objID provided by both the NED/IPAC and the Garching databases 
to match the data. In this way, we found 112 Seyfert 1s and 463 Seyfert 2s.

The  reddening correction was carried out
comparing the observed H$\alpha$/H$\beta$ ratio with the theoretical value of 2.86
\citep{1987MNRAS.224..801H}, obtained for the Case B, considering 
an electron density of 100 $\rm cm^{-3}$ and an electron temperature of
10\,000 K.  We assumed the Galactic extinction law by \cite{Miller+72} with the ratio of total to selective extinction R$_v$=3.2.
For ten objects the H$\alpha$/H$\beta$ were found to be lower than 2.86. Taking into account the errors in the measurements, for 
seven of them that present a reddening correction C(H$\beta$) between $-0.2$ and 0, we assumed it is equivalent to zero, and hence we 
did not apply any reddening correction. We take off from our sample the other three objects with C(H$\beta$) lower than $-0.2$.
The stellar mass range of our sample is $9.4 \: \la \: \log(M_{*}/ \rm M_{\odot}) \: \la \: 11.6$, somewhat wider than
the one considered by \citet{thomas19}, who found that the oxygen abundance increases by $\rm \Delta (O/H) \sim 0.1$ dex
as a function of the host galaxy stellar mass over the  $10.1 \: \la \: \log(M_{*}/ \rm M_{\odot}) \: \la \: 11.3$ range.

 The  $M_{*}$ determination of the objects in our sample is based on  a comparison between theoretical
spectra from  stellar population synthesis (SSP) codes with the SDSS  $z$-band luminosities
carried out by  \citet{2004ApJ...613..898T} and \citet{2003MNRAS.341...33K}.  The errors associated to the $M_{*}$ determinations are mainly due to
 star-formation histories, ages, metallicities and extinction assumed in the
SSPs fitting, which may differ from those of  galaxies. In general, it is
assumed the $M_{*}$ error is of the order of 0.2 dex 
(e.g. \citealt{2008A&A...488..463M, 2011MNRAS.418.1587T}).

For the resulting Seyfert 2 AGNs sample,  reddening corrected intensities
(in relation to H$\beta$=1.0) of the   
[\ion{O}{ii}]$\lambda$3726+3729, 
[\ion{Ne}{iii}]$\lambda$3869, 
[\ion{O}{iii}]$\lambda$4363, 
[\ion{O}{iii}]$\lambda$5007, 
He\,I$\lambda$5876, 
[\ion{O}{i}]$\lambda$6300, 
[\ion{N}{ii}]$\lambda$6584, 
[\ion{S}{ii}]$\lambda$6716, 
[\ion{S}{ii}]$\lambda$6731 
and [\ion{Ar}{iii}]$\lambda$7135 emission-lines, 
redshifts ($z \:\la \: 0.4)$, 
reddening correction C(H$\beta$),  the electron density (in units of particles per cm$^{3}$, see Sec.~\ref{oxy}) are listed in a Table only available in online version.
We take as zero the emission-line intensities that in the SDSS database have values lower than zero.

\subsection{Aperture effects}

The estimation of the physical properties of  objects with  different redshifts whose data were
obtained by instruments with fixed aperture, such as the  objects from the SDSS, are subject to
some degree of uncertainty. \citet{kewley05} investigated the effect of aperture size  on 
the star formation rate, $Z$ and reddening determinations for galaxies with 
distinct morphological type. Concerning the metallicity, \citet{kewley05} found that
for aperture capturing less than $20\%$ of the total galaxy emission, the derived
metallicity can differ by a factor of 0.14 dex from the value obtained when the
total galaxy emission is considered.

In our case, only properties of the nuclear region are being considered, therefore, the aperture effect
can not be so important. The diameter of the SDSS  optical fibers is $\sim 3$\arcsec, which implies that we are considering
fluxes  only emitted by the nuclear regions of the galaxies in the sample.
In fact, our sample of 463 Seyfert 2 galaxies have redshifts in the range $ 0.03 \: \la \: z \: \la \: 0.37$, 
assuming a spatially flat  cosmology with $H_{0}$\,=\,71 $ \rm km\:s^{-1} Mpc^{-1}$, $\Omega_{m}=0.270$, and $\Omega_{\rm vac}=0.730$  \citep{wright06}, 
which corresponds to   a  physical scale (D) in the center  of the disk of each galaxy
in the range $\rm 50 \: \la \: D(pc) \: \la \:   660$, i.e. the emission is mainly from the AGN.
For example, \citet{thaisa07} showed that the highest [\ion{N}{ii}]$\lambda$6584/H$\alpha$ line ratio
in the nuclear region of   NGC\,6951 (LINER/Seyfert nuclei) is  within a nuclear radius with $\sim \: 100$ pc.
 Thus,  for the farthest  objects of our sample, the measured fluxes are emitted mainly by the AGN because the flux from 
(circum)nuclear star-forming regions, if present,  have low contribution to the total flux.  The support for this assertion was found, recently, by 
\citet{thomas19}.  These authors showed that the aperture effect 
is not important on $Z$ estimations in a similar AGN
sample like the one being considered in this paper,  once similar mass-metallicity relations for
galaxies in four different redshift bins were derived in their analysis.
However, \citet{ 2018ApJ...861L...2T} pointed out that a  mixing of AGN and  \ion{H}{ii} regions emission
is expected in  the majority of  AGNs (see also \citealt{2019MNRAS.487.4153D} and reference therein).

 For the nearest objects, we could be estimating the metallicity only for the central part of
the AGNs and the metallicity of the entire AGN can be different from this little region.
Abundance  studies of spatially resolved  AGNs are (still) seldom  found in the literature and not
conclusive results have been obtained.
For example,  optical data of the nuclear region of
the Seyfert 2 galaxy Markarian 573 obtained by \citet{2018ApJ...856...46R} and \citet{thomas18a}
showed that the oxygen abundance 
is almost constant, with variations not larger than 0.10 dex along the central region.
On the other hand, \citet{thomas18a} found for two (NGC\,2992 and ESO 138-G010) of the four objects analysed 
a steep metallicity gradients from the nucleus into
the ionization cones, with $(Z/Z_{\odot})$ ranging from  $\sim 0.5$ (in the outer regions)
to $\sim 2$ (in the nucleus). 

\begin{figure} 
\includegraphics[scale=0.6, angle=-90]{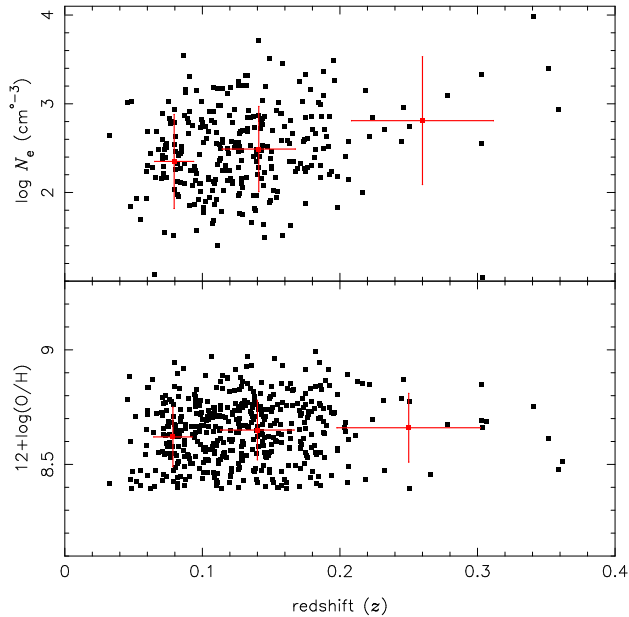}
\caption{Bottom panel: Oxygen abundance [in units of 12+log(O/H)] versus the redshift for
our sample of AGNs  (see Sect.~\ref{observ}). The oxygen abundance was calculated
using the calibration proposed by \citet{2017MNRAS.467.1507C}. Red points
represent the average and their error bars the standard deviation for the redshift bins
$z=0.0-0.1$ and $z=0.1-0.2$.
Top panel: Same than the bottom panel but for the logarithm of the electron density ($N_{\rm e}$).}
\label{oxyall}
\end{figure}

 In order to explore the presence of an aperture effect on our oxygen abundance determinations,
in the lower panel of Fig.~\ref{oxyall}, we plotted for each object of our sample the oxygen abundance values estimated using 
the calibration by \citet{2017MNRAS.467.1507C} versus the redshift, considering
 the redshift bins $z=0.0-0.1$, $z=0.1-0.2$ and
 $z \: > \: 0.2$. We calculated the average and  standard deviation of 12+log(O/H)  and $z$ for each bin.
Since it is not expected a significant chemical evolution over $z=0-0.4$, any systematic difference in the 
averages could be due to aperture effects. As can be seen in Fig.~\ref{oxyall},
the average oxygen abundances are similar for all the redshift bins  ($\approx8.64$ dex).
In the upper panel of Fig.~\ref{oxyall}, we plotted for each object of our sample the electron densities ($N_{\rm e}$) 
as a function of the redshift and the average density values, with the  standard deviations, for the same redshift bins 
definied above.
Densities were estimated from the [\ion{S}{ii}]$\lambda$6716/$\lambda$6731 emission-line ratio as described in Sec. \ref{oxy}.
Since the electron densities in AGNs are higher by about a factor of 2 than those estimated for \ion{H}{ii} regions  
(see e.g., \citealt{2000A&A...357..621C, 2003ApJ...591..801K, 
2014MNRAS.443.1291D, 2016ApJ...816...23S}), 
it is expected that if there is 
a significant contribution to the sulfur emission by \ion{H}{ii} regions in the SDSS fluxes, 
 a $N_{\rm e}$ decrement with the $z$ increases would be found.
As can be seen in the upper panel of Fig.~\ref{oxyall}, such as for the O/H, the $N_{\rm e}$ average values
in the different bins are very similar ($\approx 650 \: \rm cm^{-3}$), indicating that this
parameter does not change with the redshift.
Therefore, we assume that 
aperture effects are not significant for the parameter estimations of the objects in our sample.

\begin{figure} 
\includegraphics[scale=0.6, angle=-90]{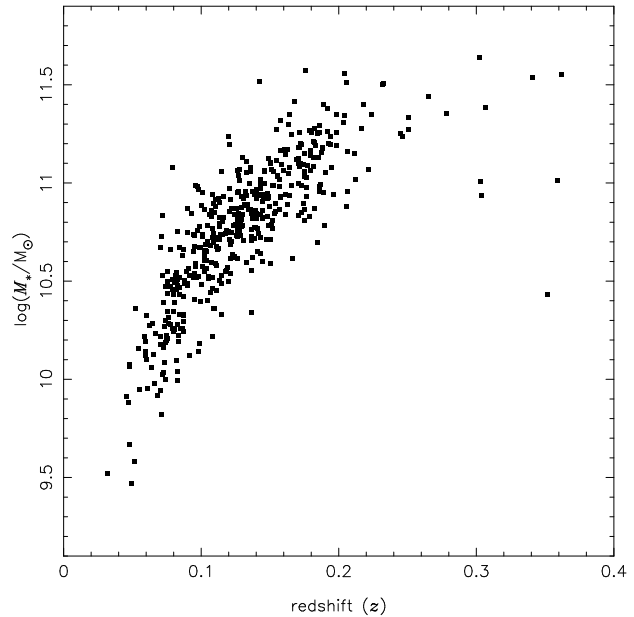}
\caption{Logarithm of the stellar mass (in units of $\rm M_{\odot}$) versus
the redshit for our objects sample (see Sect.~\ref{observ}).}
\label{massall}
\end{figure}

 Concerning the stellar mass of the galaxies in our sample, it is expected that due to aperture effects,
it increases  with $z$. This happens because as the $z$ increases, 
the projected SDSS fiber covers a larger galaxy portion and, hence, we are estimating the galaxy mass taking into account a larger area
(for a full description see, for example, \citealt{2004ApJ...613..898T}).
In order to show that, in Fig.~\ref{massall}, the logarithm of the stellar mass 
(in units of $\rm M_{\odot}$) versus the redshift for the objects in our sample is presented,
where a clear correlation is appreciated. This result indicates that to
obtain  a reliable mass-metallicity relation it is necessary to consider 
different redshift bins (see, e.g. \citealt{2008A&A...488..463M, 2008ApJ...681.1183K}).

\section{Metallicity estimations}
\label{oxy}

 We used the emission-line intensities, listed in the  online Table, to estimate the total 
 oxygen abundances relative to hydrogen  abundance (generally used as metallicity tracer)  
of the NLRs  for the objects in our sample of AGNs.  
All the methods used in this work were taken from the literature and are described  as it follows.

\subsection{$T_{\rm e}$-method}
\label{temet}

 This method consists  of calculating  the oxygen abundance in relation to
the hydrogen one (O/H) using  direct measurements of the electron temperature
of the gas phase. We  followed the methodology described in \citet{2015MNRAS.453.4102D}
which is based on \citet{2009MNRAS.398..949P}, \citet{2008MNRAS.383..209H}, \citet{2007MNRAS.381..125P}
and \citet{2003MNRAS.346..105P}.

The electron temperature in the high ionization zone of the gas phase,  referred to $t_{3}$,
for each object of the sample, was calculated from  
the observed line-intensity ratio $R_{\rm O3}$=[\ion{O}{iii}]($\lambda4959\: + \: \lambda 5007)/\lambda4363$
and using the expression 
\begin{equation}
 \label{eqt3}
 t_{3}=0.8254-0.0002415 R_{\rm O3}+\frac{47.77}{R_{\rm O3}},
 \end{equation}
where $t_{3}$ is in units of $10^{4}$ K.  This relation is valid for the
range  of $0.7  \: \la \: t_{3} \: \la \: 2.3$.

 The  electron temperature value for the low ionization zone, referred to $t_{2}$, was derived from
the theoretical relation 
\begin{eqnarray}
\label{eqo3}
t_{2}^{-1}\,=\,0.693\,t_{3}^{-1}+0.281.
\end{eqnarray}

 The electron density ($N_{\rm e}$), for each object, was calculated from the 
 $R_{S2}=$[\ion{S}{ii}]$\lambda 6716/\lambda 6731$ line ratio, 
using the {\sc IRAF/temden} task and assuming  the $t2$ value obtained from Eq.~\ref{eqo3}.
It was possible to compute $N_{\rm e}$ for 295 ($\sim 64\%$) objects of our sample.
For the other objects  $N_{\rm e}=650 \: \rm cm^{-3}$   was assumed, the average value derived for our sample.

The $\rm O^{++}$  and $\rm O^{+}$ ionic abundances in relation to $\rm H^{+}$ abundance
were computed  through the relations: 
\begin{eqnarray}
 12+\log(\frac{{\rm O^{++}}}{{\rm H^{+}}}) \!\!\!&=&\!\!\! \log \big( \frac{I(4959)+I(5007)}{I{\rm (H\beta)}}\big)+6.144  \nonumber\\
                                          &&\!\!\!+\frac{1.251}{t_{3}}-0.55\log  t_{3}   
   \end{eqnarray}
   and
\begin{eqnarray}
 12+\log(\frac{{\rm O^{+}}}{{\rm H^{+}}})  \!\!\!&=&\!\!\! \log  \big( \frac{I(3727)}{I{\rm (H\beta)}}\big)+5.992 \nonumber\\
                                           &&\!\!\!+\frac{1.583}{t_{2}}-0.681\log t_{2} +\log(1+2.3 n_{\rm e}),
\end{eqnarray}
where $n_{\rm e}$ is the electron density $N_{\rm e}$ in  units of 10\,000 $\rm cm^{-3}$.

Finally, the total oxygen abundance in relation to hydrogen one (O/H) was calculated assuming
\begin{eqnarray}
\rm
\frac{O}{H}=\frac{O^{+}}{H^{+}}+\frac{O^{++}}{H^{+}}.
\label{icf1}
\end{eqnarray}
 The expression above  assumes that the Ionization Correction Factor (ICF) for the oxygen is equal to 1,
even though  ions with higher ionization states are observed in other spectral bands as, for instance, 
X-rays (e.g. \citealt{2009A&A...505..541C, 2011A&A...530A.125C, 2010MNRAS.405..553B,
2017ApJ...848...61B}), indicating that there could be a significant contribution of them. 
We point out this issue in the work by \citet{2019MNRAS.489.2652P}. A model-base estimation 
of the oxygen ICF for NLRs will be addressed in a forthcoming work even though in Sec.\ \ref{disc} 
we provided a brief review of alternative ICF(O) values. 

Due to the fact that the   [\ion{O}{iii}]$\lambda$4363 line is weak or not observable in the majority
of AGNs and, due to the validity range of the Equation~\ref{eqt3},
it was possible to apply the $T_{e}$-method only in  154 ($\sim 33 \%$)  objects of our sample.

\subsection{Strong-line method}

\subsubsection{Storchi-Bergmann et al. calibrations}

\citet{1998AJ....115..909S} proposed the first calibrations between the metallicity [$Z\rm=12+\log(O/H)$]
and the intensities of 
narrow optical emission-line ratios of AGNs. These calibrations are    based on
results of photoionization models built with the {\sc Cloudy} code.
The calibrations proposed by these authors are:

\begin{eqnarray}
       \begin{array}{l@{}l@{}l}
\rm (O/H)_{SB98,1} & = &  8.34  + (0.212 \, x) - (0.012 \,  x^{2}) - (0.002 \,  y)  \\  
         & + & (0.007 \, xy) - (0.002  \, x^{2}y) +(6.52 \times 10^{-4} \, y^{2}) \\  
         & + & (2.27 \times 10^{-4} \, xy^{2}) + (8.87 \times 10^{-5} \, x^{2}y^{2}),   \\
     \end{array}
\label{sb1}
\end{eqnarray}

\noindent where $x$ = [N\,{\sc ii}]$\lambda$$\lambda$6548,6584/H$\alpha$ and 
$y$ = [O\,{\sc iii}]$\lambda$$\lambda$4959,5007/H$\beta$ and

\begin{eqnarray}
       \begin{array}{lll}
(\rm {O/H})_{{\rm SB98,2}}   & = &  8.643 - 0.275 \, u + 0.164 \, u^{2}   \\  
          & + & 0.655 \, v - 0.154 \, u v  - 0.021 \, u^{2}v \\  
          & + & 0.288 v^{2} + 0.162 u v^{2} + 0.0353 u^{2}v^{2},   \\
     \end{array}
\label{sb2}
\end{eqnarray}

\noindent where $u$ = log([O\,{\sc ii}]$\lambda$$\lambda$3727,3729/[O\,{\sc iii}]$\lambda$$\lambda$4959,5007) 
and $v$ = log([N\,{\sc ii}]$\lambda$$\lambda$6548,6584/H$\alpha$).  The term O/H above corresponds
to 12+log(O/H). Both calibrations are valid for   $\rm 8.4 \: \lid \: 12+log(O/H) \:  \lid \: 9.4$ and were
obtained adoptating in the models the (N/O)-(O/H) abundance relation derived for nuclear starbursts
by \citet{1994ApJ...429..572S}.

As pointed out by \cite{1998AJ....115..909S}, the  O/H should be
corrected  in order to take into account the electron density ($N_{\rm e}$)
effects. Hence, the final value for the ratio O/H ratio is given by  the
relation below:

\begin{equation}
{\rm (O/H)_{final}=[(O/H)}-0.1 \: \times \: \log (N_{\rm e}/300 ({\rm cm^{-1}})].
\end{equation}

\subsubsection{Castro  et al. calibration}
 
\citet{2017MNRAS.467.1507C} proposed a semi-empirical calibration between the 
metallicity $Z$ and the line ratio  $N2O2$=log([\ion{N}{ii}]$\lambda$6584/[\ion{O}{ii}]$\lambda$3727).
This calibration   was performed determining  $Z$ of a sample of  58   Seyfert 2 AGNs through a
diagram containing the observational data and the results of a grid of photoionization models obtained with the 
{\sc Cloudy} code \citep{2013RMxAA..49..137F}. In  these models, the (N/O)-(O/H) abundance relation derived for  
\ion{H}{ii} regions by \citet{2000ApJ...542..224D} was assumed. These authors found
 \begin{eqnarray}
     \begin{array}{lll}
(Z/Z_{\odot}) \!\!\!  & =  &\!\!\! 1.08(\pm0.19) \times N2O2^2  +  1.78(\pm0.07) \times N2O2  \\
                      &    &           +1.24(\pm0.01) . \\  
     \end{array}
\label{eq3}
\end{eqnarray}
The oxygen abundance  is obtained by
\begin{equation}
12+\log({\rm O/H})=12+\log[(Z/Z_{\odot})\times 10^{-3.31}],
\end{equation}
where the solar oxygen abundance $\rm \log(O/H)=-3.31$ derived by 
\citet{allendeprieto01} was considered.

\subsubsection{{\sc \ion{H}{ii}-Chi-mistry} code}

The {\sc \ion{H}{ii}-Chi-mistry} code  (hereafter {\sc HCm}), proposed by  \citet{enrique14}, establishes a bayesian-like comparison 
between the predictions from a grid of photoionization models covering a large range of input parameters
and  using the lines emitted by the
ionized gas.  This method has the advantage of not assuming any fixed relation between  secondary  and primary elements 
(e.g. N-O relation) considered in most of the calibrations such as the ones proposed by \citet{1998AJ....115..909S} and \citet{2017MNRAS.467.1507C}.
In \citet{2019MNRAS.489.2652P} this code was adapted to be used in the Seyfert 2 AGNs and  this last version is the one considered here.

 \citet{thomas19} proposed another Bayesian code ({\sc NebulaBayes}) presented initially by \citet{blanc15}
and  based on a comparison 
between observed emission-line fluxes  and photoionization model grids
which helped to obtain robust  measurements of
abundances in the extended narrow-line regions (ENLRs) of AGNs.
This code produces very similar O/H values to those found using the 
calibration of \citet{2017MNRAS.467.1507C}, such as pointed out by 
 \citet{thomas19}. Therefore, by  simplicity, we do not consider it here.

 The oxygen abundance estimations for each object of the sample computed by using the methods above
are listed in the online Table.

\section{Results}
\label{res}

\begin{figure*}
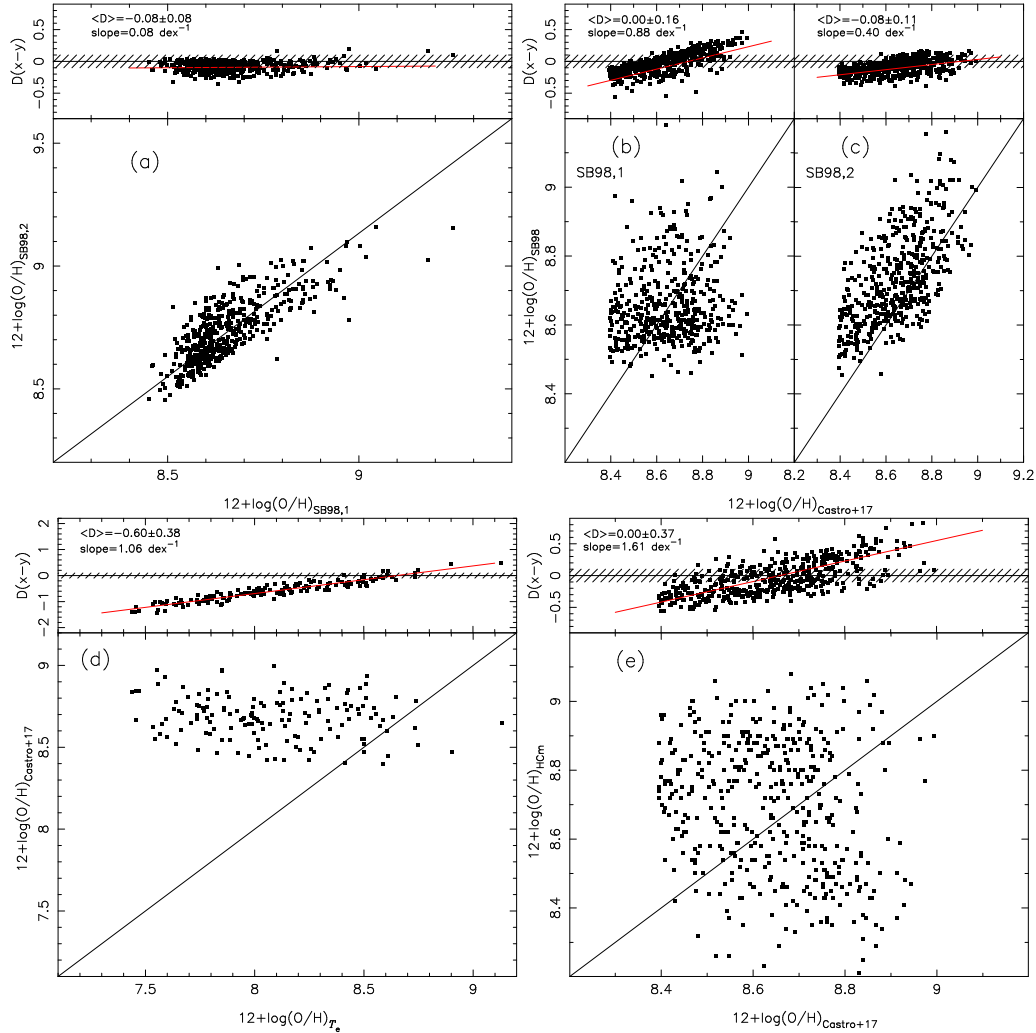

\centering
\includegraphics[scale=0.5, angle=-90]{comp_thaisa.eps}
\includegraphics[scale=0.5, angle=-90]{comp.eps}\\
\includegraphics[scale=0.5, angle=-90]{comp2.eps} 
\includegraphics[scale=0.5, angle=-90]{comp3.eps}\\
\caption{Comparison between  oxygen abundances
[in units of 12+log(O/H)] computed using the observational data described in Sect.~\ref{observ}
and the methods listed in Sect.~\ref{oxy}. Bottom panel of each plot is the comparison
between two estimations. Solid line represents the equality between 
these. Top panel is the difference (D=x-y) between the  estimations. Black line
represents the null difference while red  line represents a linear regression to these differences whose
slope is indicated.
The average difference ($\rm <D>$) is indicated in each plot. The dashed area indicates the
uncertainty of $\pm 0.1$ assumed in the oxygen abundance estimations.
Panel (a): Comparison between  oxygen abundances computed via Eq.s~\ref{sb2} versus Eqs.~\ref{sb1} proposed by \citet{1998AJ....115..909S}.
Panel (b): Such as panel (a) but for Eq.~\ref{sb2} (referred to SB98,1) versus O/H estimations via
\citet{2017MNRAS.467.1507C} calibration. Panel (c): Such as panel (a) but for Eq.~\ref{sb1}  (referred to SB98,2) versus O/H estimations via
\citet{2017MNRAS.467.1507C} calibration. Panel (d): Such as panel (a) but for O/H estimations via
\citet{2017MNRAS.467.1507C} versus the ones via $T_{\rm e}$-method. Panel (e): Such as panel (a) but for O/H estimations via
{\sc Hii-Chi-mistry} ({\sc HCm}) code versus the ones via \citet{2017MNRAS.467.1507C}  calibration. }
\label{fig2}
\end{figure*}

We used the observational data described in Sect.~\ref{observ} in order to compare the oxygen abundance
estimations computed  using the  aforementioned methods. 

For SFs, the metallicity or oxygen abundance   is defined by  estimations
based on the classical $T_{\rm e}$-method
and any calibration must be tested
comparing its estimations to this {\it bona fide}  method.
 The accuracy of the $T_{\rm e}$-method is also supported  by the agreement between
oxygen abundances in nebulae located in the  solar neighborhood and those derived from observations
of the weak interstellar \ion{O}{i}$\lambda$1356 line towards the stars (see \citealt{pilyugon03} 
and references therein),  although determinations of stellar oxygen abundances following different approaches have led
to distinct values with variations of up to $\sim 3$ dex, as showed by \citet{2015A&A...579A..88C}.
However, as pointed out by \citet{2015MNRAS.453.4102D}, the $T_{\rm e}$-method, in its usual application form, does not work for AGNs and, obviously, 
it can not be used as reference for this kind of object. In other words, there is  no consensus on which is the best method to 
estimate O/H (or $Z$) in AGNs. Therefore, we compared O/H estimations based on the methods listed above to know the discrepancy 
between them. 

The uncertainty in the metallicity estimations (traced by the O/H abundance) 
 depends on which method is considered. For example, for the $T_{\rm e}$-method the uncertainty is about
0.1 dex (e.g. \citealt{leonid00, 2003ApJ...591..801K, 2008MNRAS.383..209H}) while for  strong-line calibrations 
is in order of 0.2 dex \citep{denicilo02}.  In this paper, we assume that the uncertainty in  O/H estimations
is 0.2 dex, the highest uncertainty value considered in \ion{H}{ii} region abundance studies.

We start the analysis comparing the O/H estimations computed from the two \citet{1998AJ....115..909S} calibrations
(SB98,1 and SB98,2).
In Fig.~\ref{fig2}, panel (a), the oxygen abundances  calculated using SB98,2 versus SB98,1 
 are shown.  A good agreement  between the estimations can be seen, with SB98,1 producing 
somewhat lower values ($-0.08$ dex) than the ones from SB98,2. 
\citet{1998AJ....115..909S} carried out a similar comparison
but using  only 7 objects and these authors found differences of about $-0.1$ dex, about the same value
derived by us. More recently, \citet{2015MNRAS.453.4102D}  
compared O/H values predicted by photoionization models with estimations obtained from \citet{1998AJ....115..909S} calibrations, 
in an O/H versus $R_{23}$=([\ion{O}{ii}]$\lambda$3727+[\ion{O}{iii}]$\lambda$4959+$\lambda$5007)/H$\beta$ plot,
and these authors found a better  consistency with the SB98,1.
\citet{2015MNRAS.453.4102D} used a small sample (47 Seyfert 2s). However, taking into account
the uncertainty of 0.2 dex, both \citet{1998AJ....115..909S} calibrations produce similar abundances.

In Fig.~\ref{fig2}, panels (b) and (c), we compare the estimations via the two \citet{1998AJ....115..909S} calibrations
with the ones obtained via \citet{2017MNRAS.467.1507C} calibration. We can see that, despite the difference between 
the estimations, the  average difference is lower than the uncertainty.
However,  a  systematic discrepancy is clearly derived, in the sense
that \citet{1998AJ....115..909S} calibrations produce lower and higher values for the high  ($\rm 12+\log(O/H)\: \ga \: 8.6$) 
and low  ($\rm 12+\log(O/H)\: \la \:8.6$)  metallicity regimes,
respectively,  being this behavior more clear when  the SB98,1 is considered.
One can note  that the difference between estimations from  \cite{2017MNRAS.467.1507C} and 
from \citet{1998AJ....115..909S} calibrations reach up to 0.5 dex
for the lowest metallicity values ($\rm 12+\log(O/H)\approx7.5$). \citet{2017MNRAS.467.1507C} found a similar result, 
although most of the objects considered by these authors are located
around 12+log(O/H)=8.7, i.e. the solar abundance \citep{allendeprieto01}.
In Fig.~\ref{fig2}, panel (d), the estimations by the calibration by \citet{2017MNRAS.467.1507C}
versus those obtained via $T_{\rm e}$-method are shown. A  systematic difference  is found, 
ranging from $\sim 0$ for the highest O/H values to $\sim 2$ dex for the lowest ones.
The average difference is about $-0.6$ dex, a lower value than the one ($-0.8$ dex) found by \citet{2015MNRAS.453.4102D},
 who compared O/H estimations derived using  \citet{1998AJ....115..909S} calibrations with those via $T_{\rm e}$-method.
 In Fig.~\ref{fig2}, panel (e), the values derived from \citet{2017MNRAS.467.1507C} are compared  to those from 
{\sc HCm} code (P\'erez-Montero et al. 2019), where, despite the difference between estimations is about zero,
a systematic difference is found.

\section{Discussion}
\label{disc}
It is known that in SFs many strong-line methods calibrated using theoretical models overestimate $Z$
as compared  to the results from the $T_{\rm e}$-method.
For example, \citet{2007A&A...462..535Y}
determined the gas-phase oxygen abundance using the  $T_{\rm e}$-method for a sample of 695 star-forming galaxies and \ion{H}{ii} regions with reliable detections of 
[\ion{O}{iii}]$\lambda$4363. These authors found that the oxygen abundances derived using certain  theoretical calibrations are between  0.06 and 0.20 tex  larger
 than those derived using the $T_{\rm e}$-method.
\citet{2008ApJ...681.1183K} analyzed the mass-metallicity (\textit{M--Z}) relation of star-forming galaxies, 
whose data were taken from the SDSS \citep{2000AJ....120.1579Y} database, and found metallicity discrepancies  for a 
fixed value of $M$  of up to $\sim0.7$ dex when  distinct theoretical and empirical strong-line methods are considered.
 
Regarding AGNs, when only strong-line methods are considered, discrepancies  of up to $\sim0.8$ dex were  found when  distinct methods are used to 
estimate O/H in NLRs of Seyfert 2s,  being these discrepancies higher for the low metallicity regime ($\rm 12+\log(O/H) \: \la \: 8.5$).
The discrepancy found when the \citet{1998AJ....115..909S} and
\citet{2017MNRAS.467.1507C} calibrations are considered are due to the different  N-O abundance relations
assumed  in the photoionization models by these authors, which  have a more important effect for the
low metallicity regime, mainly because [\ion{N}{ii}] lines are used in both calibrations
\citep{2009MNRAS.398..949P}. The discrepancy between the  $Z-N2O2$ calibration \citep{2017MNRAS.467.1507C}
and those derived from the bayesian code {\sc  HCm}  \citep{2019MNRAS.489.2652P}
can also be due to  a  fixed N-O relation. In fact,
as mentioned, \citet{2017MNRAS.467.1507C} assumed photoionoization models with fixed  N-O relation, taken from \ion{H}{ii}
chemical abundance estimations carried out by \citet{2000ApJ...542..224D}, while in the bayesian {\sc HCm} approach this relation is not fixed.

 The $T_{\rm e}$-method produces, possibly,  unreal O/H subsolar 
estimations and the origin of these low values could arise from the supposition that the ICF for the oxygen is equal to 1 (Eq.\ \ref{icf1}). 
There are no equations to estimate oxygen ICFs for NLRs of type 2 AGNs in the literature. 
For Planetary Nebula (PN), the following expression to estimate ICF(O) was proposed by \citet{torrespeimbert77}:   
\begin{equation}
\label{icfox}
\rm
ICF(O)=\frac{N(He^{+}+He^{2+})}{N(He^{+})},
\end{equation}
where N represents the abundance \citep[see also][]{alexander97,izotov06, jorge07, delgadoinglada14}. 
 This equation provides estimated values for the ICF of PNs in the range between 
 $\sim 1$ and 1.6 \citep[e.g.][]{krabbe06}.
For \ion{H}{ii} regions, low ICF(O) has been also derived (e.g.\ \citealt{izotov06}). 
Unfortunately, for  the objects in our sample it was not possible
to apply Eq.~\ref{icfox} because the \ion{He}{ii}$\lambda$4686 emission line, 
necessary to calculate $\rm N(He^{2+})$, was not measured. For this reason, we used
the  sample of 47 type 2 AGNs   compiled by \citet{2015MNRAS.453.4102D} in order to
calculate the ICF(O). We used the expressions by \citet{izotov94}:
\begin{equation}
\label{abundhe}
 {\rm \frac{N(He^{+})}{N(H^{+})}}=0.738 \: t^{0.23} \: \frac{I(\lambda 5876)}{I(\rm H\beta)}
\end{equation}
and
\begin{equation}
 {\rm \frac{N(He^{2+})}{N(H^{+})}}=0.084 \: t^{0.14} \: \frac{I(\lambda 4686)}{I(\rm H\beta)},
\end{equation}
where  $t=t_{3}$  is assumed. It was possible to calculate the ICF(O) only for 33 objects
since the \ion{He}{i}$\lambda$5876
and \ion{He}{ii}$\lambda$4686  emission lines are not available for all these 47 objects.

\begin{figure} 
\includegraphics[scale=0.6, angle=-90]{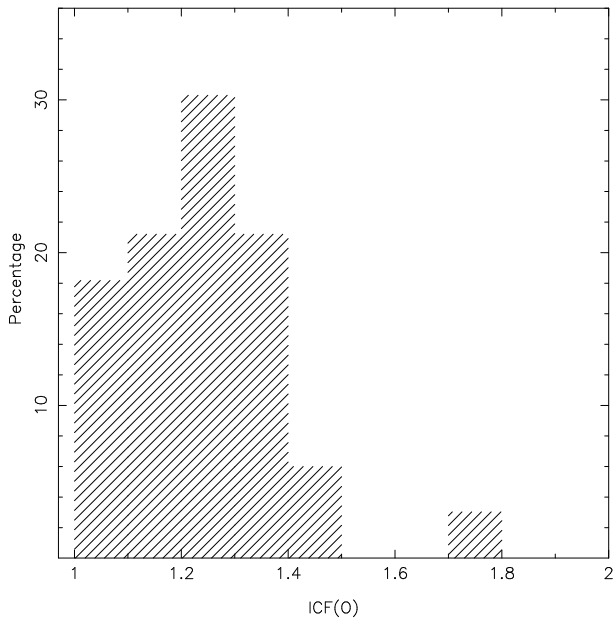}
\caption{Histogram containing the ICF(O) calculated using the
Eqs.~\ref{icfox} and \ref{abundhe} for 33 type AGNs whose data were
compiled by \citet{2015MNRAS.453.4102D}.}
\label{hist1}
\end{figure}

In  Fig.~\ref{hist1} a histogram with the ICF(O) distribution is shown.
It can be seen that most part of the objects have $\rm ICF(O) \: \la \: 1.4$,  
with an average value of $1.23 \pm0.15$. This indicates an average  oxygen abundance correction of about 
0.1 dex, i.e. the oxygen in AGNs is  mainly in $\rm O^{+}$ and $\rm O^{++}$ ionic stages.
Therefore, the  supposition of ICF(O)=1 would not  be
the cause of the discrepancy derived between O/H estimations based on $T_{\rm e}$-method
and on strong-line methods.  It must be noted that, 
as we pointed above, we are using an ICF(O) derived for Planetary Nebulae.

\begin{figure} 
\includegraphics[scale=0.6, angle=-90]{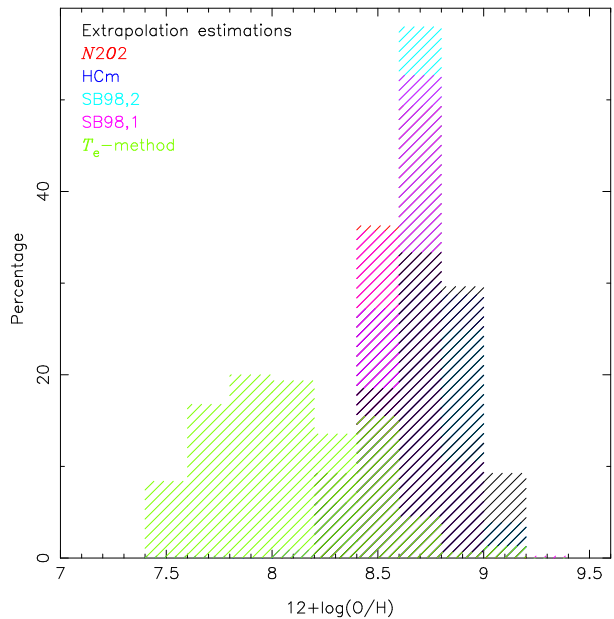}
\caption{Histogram containing the oxygen abundance distributions for
 NLRs of AGNs and based on distinct methods as indicated and described in Sect.~\ref{oxy}. 
Extrapolation estimations refers to the extrapolated values to the nuclear region obtained using the radial oxygen gradient 
derived by \citet{2004A&A...425..849P}  for a sample of spiral galaxies.}
\label{hist}
\end{figure}

In Fig.~\ref{hist}   we show the histograms of the  oxygen abundances 
in the selected sample derived following the
distinct methods described in  Sect.~\ref{oxy}, as
compared with the abundances obtained
by  extrapolating the O/H radial distributions to the nuclear region (containing AGN and SF region) of a sample of spiral disks
obtained by \citet{2004A&A...425..849P}, who used the $P$-method
\citep{2001A&A...369..594P}.  These extrapolated estimations can be understood as an independent
ones, which do not suffer effects of intrinsic uncertainties present
in photoionization models or the limitations of the $T_{\rm e}$-method.
 We can see that  strong-line methods produce similar
oxygen abundance distributions, with the most frequent value around of 8.7,
the solar abundance.   On the other hand, the $T_{\rm e}$-method
produces, in most cases, sub-solar abundances. 
We list in Table~\ref{tab1}  the minimum, maximum and
average values of the distributions of oxygen abundances derived   using the distinct methods described in
Sect.~\ref{oxy}. From the above results one can conclude that, considering the uncertainty 
of 0.2 dex in the oxygen estimations, all strong-line methods available in the literature produce
similar oxygen abundance distributions when a large and   homogeneous sample
of data are used.  
The average maximum value of the oxygen abundance  for 
our sample of Seyfert 2 AGNs through the strong-line methods is $\rm 12 + \log(O/H)\sim9.1$, which is 
slightly higher than the one derived for star-forming galaxies ($\sim8.95$ dex) by \citet{2007MNRAS.376..353P}.
This agreement suggests that there is no extraordinary chemical enrichment of the NLRs of AGNs (see also \citealt{2015MNRAS.453.4102D}), 
as also pointed out from the comparison between N and O abundances both in AGNs and in SFs by  \citet{2017MNRAS.468L.113D} and
\citet{2019MNRAS.489.2652P}.

\begin{figure*} 
\includegraphics[scale=1.0, angle=-90]{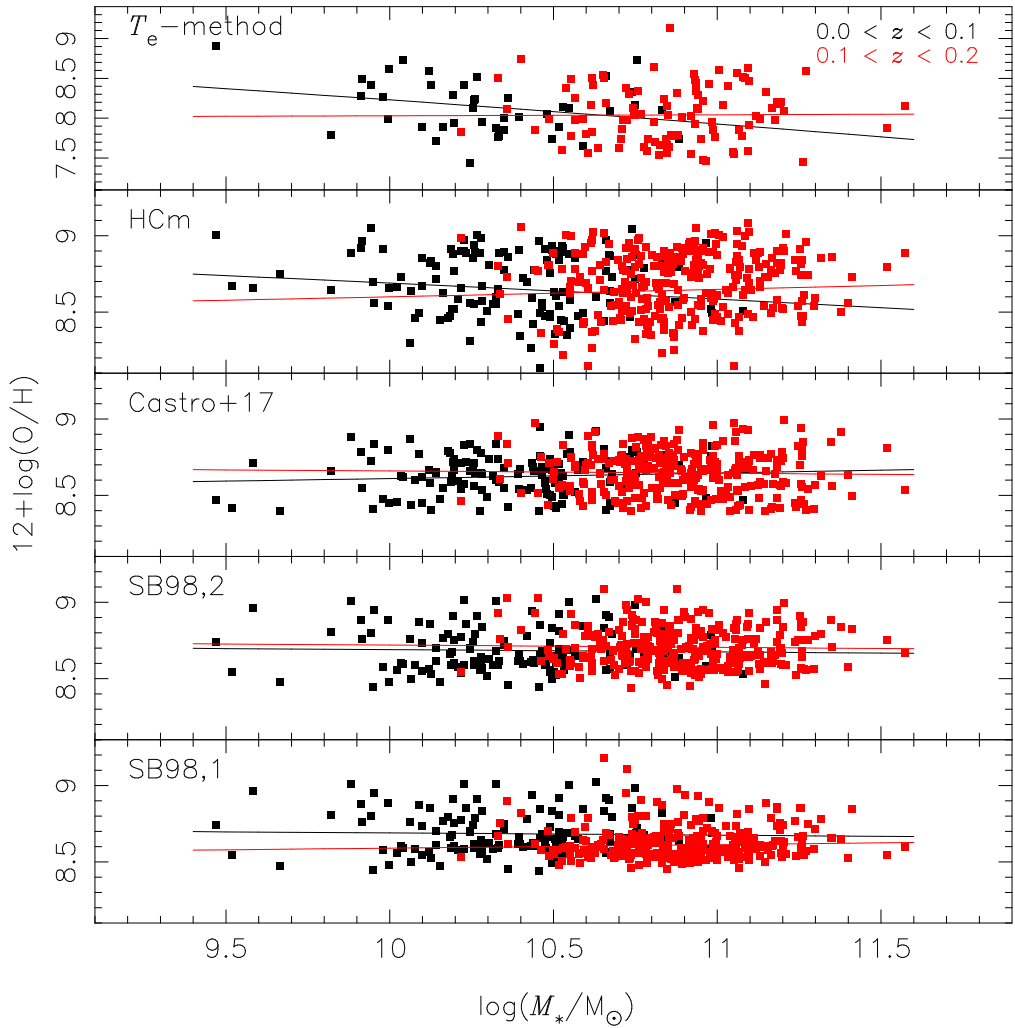}
\caption{Metallicity of the NLR for the sample of  AGNs (see Sect.~\ref{observ}) versus the logarithm of 
stellar mass $M_{*}$ (in units of solar mass ($\rm M_{\odot}$) of the hosting galaxy.
The methods considered to obtain the $Z$ estimations are indicated in each plot. The curves represent
the fitting of the Eq.~\ref{mzr} on the points whose coefficients are listed in  Table~\ref{tab1}.
Different color correspond to estimations for two redshift intervals indicated in the top panel.}
\label{fig3}
\end{figure*}

\begin{table*}
\caption{Minimum, maximum and the average oxygen abundance values derived by 
the use of the  distintive methods descrived in Sect.~\ref{oxy} and indicated
in the first column. The $A$	and $B$  values correspond to the
parameter fittings of the Eq.~\ref{mzr} on the estimations showed in Fig.~\ref{fig3}
 for the redshift bins $z=0.0-0.1$ and $z=0.1-0.2$.}
\label{tab1}
\begin{tabular}{lccccccccc}
\hline
                   &\multicolumn{3}{c}{12+log(O/H)}&  &\multicolumn{2}{c}{$z=0.0-0.1$}    &  &\multicolumn{2}{c}{$z=0.1-0.2$}\\
\cline{2-4}
\cline{6-7}
\cline{9-10}
  Method           & Min.  & Max.  &  Aver. 	   &  &   $A$                &  $B$           &  &   $A$                             & $B$ \\
\hline              
 $N2O2$            & 8.39  & 8.99 & 8.64$\pm0.13$  &  & $0.0017\pm0.0019$    &$8.44 \pm0.20$  &  &$-0.0007\pm 0.0015$     & $8.73 \pm 0.18$    \\
{\sc HCm}          & 7.17  & 9.08 & 8.71$\pm0.30$  &  & $-0.0050\pm0.0004$   &$9.20 \pm0.47$  &  &$ 0.0024\pm 0.0034$     & $8.35 \pm 0.41$     \\ 
SB98,1             & 8.43  & 9.18 & 8.61$\pm0.11$  &  & $-0.0007\pm0.0015$   &$8.69 \pm0.17$  &  &$ 0.0007 \pm 0.0013$    & $8.53 \pm  0.16$    \\
SB98,2             & 8.42  & 9.18 & 8.69$\pm0.13$  &  & $-0.0005\pm0.0001$   &$8.74 \pm0.21$  &  &$-0.0011 \pm 0.0015$    & $8.83 \pm  0.18$     \\
$T_{\rm e}$-method & 7.43  & 9.13 & 8.07$\pm0.34$  &  & $-0.0144\pm0.0081$   &$9.67 \pm0.86$  &  & $0.0006 \pm 0.0006$    & $7.97 \pm  0.79$     \\
\hline
\end{tabular}
\end{table*}

We also derive the relation between the stellar mass ($M_{*}$) of the host galaxy with 
the metallicity $Z$ of its  AGN derived using the different methods analised in the present work. 
Recently, \citet{thomas19} found a mass metallicity (\textit{M--Z}) relation for 
Seyfert 2 galaxies in the local Universe   ($z \: \la 0.2$) while \citet{2018A&A...616L...4M} found this relation for type 2 AGNs 
at $1.2\:  < \: z \: < \: 4.0$ (see also \citealt{2019MNRAS.486.5853D}). 
The different \textit{M--Z}  relations are shown in Fig.~\ref{fig3}.
For each \textit{M--Z} plot we fit the expression
\begin{equation}
\label{mzr}
 12+\log({\rm O/H}) = A \times    \left(\log \frac{M_\star}{M_{\odot}} \right )^{2}    + B 
\end{equation}
 which was adapted from \citet{2008A&A...488..463M}, who derived the \textit{M--Z} relation for galaxies at different redshifts. 
The results of the fits are shown in Table~\ref{tab1} and presented in Fig.~\ref{fig3}.
 We can see that the chemical abundances derived using strong-line methods do not 
show any correlation between the metallicity of the NLR and the stellar mass of the host galaxy.

\section{Conclusion}
\label{conc}

 We used observational emission line intensities of 463 confirmed AGNs  
taken from the SDSS DR7, whose classification as Seyfert 2 is available in the NED, to compare  oxygen abundance in the
NLRs of these objects obtained through the strong-line methods: two theoretical calibrations
proposed by \citet{1998AJ....115..909S}, the semi-empirical $N2O2$ calibration proposed by \citet{2017MNRAS.467.1507C}, 
the bayesian {\sc \ion{H}{ii}-Chi-mistry} ({\sc HCm}) code proposed
by \citet{2019MNRAS.489.2652P}, as well as O/H values obtained 
by using the $T_{\rm e}$-method.   We found that the two calibrations of \citet{1998AJ....115..909S}
produce very similar oxygen abundance values from each other, with an average difference of 0.08 dex,   
a  lower value than the one (0.2 dex) attributed to uncertainty in estimations via strong-line methods.
The \citet{1998AJ....115..909S} calibrations and the {\sc HCm} code produce lower and higher O/H values 
for the high  ($\rm 12+\log(O/H)\: \ga \: 8.6$) and low  ($\rm 12+\log(O/H)\: \la \:8.6$)  metallicity regimes
in comparison to those derived by using the $N2O2$ calibration. These discrepancies are due to the relation between the nitrogen
and oxygen abundances assumed in the photoionization models considered in the calibrations (methods). 
A sistematic difference between O/H values calculated
via $T_{\rm e}$-method and via $N2O2$ calibration was found, ranging from  $\sim 0$ for the highest 
O/H values to $\sim 2$ dex for the lowest ones. We showed that this difference can not be explained 
by taking into account the use of Ionization Correction Factors for the oxygen in the $T_{\rm e}$-method.
We also analysed the influence  of the
use of the  different strong-line methods on the derivation of the relation between the stellar mass 
of the galaxies ($M_{*}$) and  the metallicity $Z$ (traced by the O/H abundance) of their AGNs.
We did not find any correlation between $Z$ and $M_{*}$ and this result is independent of the 
method used to estimate the metallicity.

\section{Acknowledgements}
We thank the referee for helping to improve this paper with her/his constructive feedback. OLD and ACK thank FAPESP and CNPq for the finnancial
support. Funding for the SDSS and SDSS-II has been provided by the Alfred P. Sloan Foundation,
the Participating Institutions, the National Science Foundation, the U.S. Department of Energy,
the National Aeronautics and Space Administration, the Japanese Monbukagakusho, the Max Planck
Society, and the Higher Education Funding Council for England. The SDSS Web Site is 
http://www.sdss.org/. The SDSS is managed by the Astrophysical Research Consortium 
for the Participating Institutions. The Participating Institutions are the American 
Museum of Natural History, Astrophysical Institute Potsdam, University of Basel, 
University of Cambridge, Case Western Reserve University, University of Chicago, 
Drexel University, Fermilab, the Institute for Advanced Study, the Japan Participation Group,
Johns Hopkins University, the Joint Institute for Nuclear Astrophysics, the Kavli Institute
 for Particle Astrophysics and Cosmology, the Korean Scientist Group, the Chinese Academy 
of Sciences (LAMOST), Los Alamos National Laboratory, the Max-Planck-Institute for 
Astronomy (MPIA), the Max-Planck-Institute for Astrophysics (MPA), 
New Mexico State University, Ohio State University, 
University of Pittsburgh, University of Portsmouth, Princeton University, 
the United States Naval Observatory, and the University of Washington.
We also thank to Max-Planck-Institute for Astrophysics and John Hopkins 
University for Physical properties for galaxies and active galactic nuclei 
in the Sloan Digital Sky Survey: Data catalogues from SDSS studies at MPA/JHU.
This research has made use of the VizieR catalogue access tool, 
CDS, Strasbourg, France (DOI : 10.26093/cds/vizier). 
The original description of the VizieR service was published 
in A$A$S 143, 23. This work has also made use of the computing 
facilities of the Laboratory of Astroinformatics (IAG/USP, NAT/Unicsul),
whose purchase was made possible by the Brazilian agency FAPESP (grant 2009/54006-4) and the INCT-A.

\bibliographystyle{mn2e}
\bibliography{publicacao}

\label{lastpage}
\end{document}